# Intrinsic network dynamics catalyze switching in functional connectivity states of the brain

Steven S. Kang[1*] and Dae-Jin Kim[2]
[1]University of Missouri-Kansas City, MO, USA
[2]Indiana University, IN, USA
*kangseung@umkc.edu

## ABSTRACT

Dynamic functional connectivity (dFC) reflects ongoing reconfiguration of large-scale brain networks, but the temporal organization and network processes underlying spontaneous state transitions remain poorly understood. We developed functional connectivity state switching (FCSS), a phase-coherence-based framework that tracks multiscale changes in dominant whole-brain connectivity configurations. In the Human Connectome Project fMRI, FCSS was strongly evoked by cognitive state and condition switches, and distinct transition dynamics differentially predicted working-memory performance. During rest, spontaneous FCSS was characterized by progressive changes in network segregation, integration, and cross-network communication as connectivity approached a switching event. Critically, event-related analysis time-locked to spontaneous FCSS revealed opposing pre-switch dynamics: salience-network contribution increased while default-mode-network contribution decreased, suggesting coordinated engagement and disengagement of these networks before switching. This reciprocal reweighting may provide a network-level mechanism for spontaneous FC reconfiguration. Together, these findings identify both global reorganization and network-specific processes underlying spontaneous and task-evoked brain-state transitions in the human brain.

Functional organization of the brain dynamically varies across time and cognitive states. Although static functional connectivity (sFC) has revealed stable and reproducible patterns of large-scale network organization, dynamic functional connectivity (dFC) captures temporal fluctuations in these interactions. Since around 2010, diverse dFC approaches using functional magnetic resonance imaging (fMRI) have been developed to characterize spontaneous and task-related changes in functional brain networks[1]. Rest and task states largely share an intrinsic network architecture, while task engagement produces task-general and task-specific modifications of this organization[2]. Time-varying network properties, including network integration, flexibility, and state occupancy and transitions, have also been linked to cognitive performance[3-5], supporting the behavioral relevance of dynamic network organization. During rest, spontaneous FC fluctuations form recurring network configurations[6], and large-scale network transitions have been linked to individual traits and cognitive-state boundaries[7]. Interactions among the salience, executive and default-mode networks have further been implicated in switching between large-scale brain states[8]. However, how spontaneous and task-evoked FC transitions differ, how the fine-grained temporal evolution of FC reconfiguration around cognitive state and condition switches relates to behavioral performance, and what network processes accompany spontaneous FC switching remain poorly understood.

This knowledge gap is partly attributable to limitations of existing dFC measures, particularly in resolving the precise temporal evolution of FC reconfiguration. Sliding-window approaches remain dominant in the field, estimating

FC from correlations within short segments of blood oxygen level dependent (BOLD) time series[1,9]. Although temporal averaging improves estimation stability, it reduces temporal resolution and may blur brief connectivity changes; results also depend on choices of window length and step size, whereas shorter windows increase sampling variability and spurious fluctuations[9,10]. Higher-temporal-resolution approaches, including multiplication of temporal derivatives, edge time series (ETS), and instantaneous phase-synchronization measures, reduce reliance on conventional correlation windows[11-13]. However, derivative-based measures often require temporal smoothing, ETS reflects framewise products of BOLD amplitudes and is dominated by high-amplitude cofluctuations, and instantaneous measures remain susceptible to single-frame variability. Moreover, these approaches yield high-dimensional pairwise connectivity patterns but do not themselves provide a robust measure of the continuous temporal trajectory of whole-brain FC reconfiguration. Other approaches define discrete FC states using clustering or hidden Markov models and identify transitions when the inferred state changes[6,14-16], whereas change-point methods identify significant differences between FC configurations before and after candidate time points[17]. Although rigorous, these approaches discretize continuous network dynamics and may obscure gradual, multistage, or individual-specific reconfiguration; inferred states and transition points can also depend on assumptions regarding state number, model selection, or significance criteria.

To address these limitations, we developed a dFC framework to characterize the temporal evolution of brain network state transitions. Time-resolved FC was estimated using instantaneous phase coherence (IPC)[12], and a dFC state was represented by the leading eigenvector (V1) of the IPC matrix[18,19], capturing its dominant whole-brain configuration while reducing dimensionality and the influence of spurious pairwise fluctuations. Multiscale comparison of adjacent V1 configurations generated a functional connectivity dissimilarity (FCD) time series, revealing gradual changes in dominant network configurations punctuated by transient, prominent peaks, which we termed functional connectivity state switching (FCSS). Using Human Connectome Project resting-state and working-memory fMRI, we examined FCD across rest and task, responses to state and condition switches, and their relationships with task performance. Resting-state analyses characterized state-dependent and trait-like properties of spontaneous FCSS. Graph-theoretical and event-related dFC analyses further identified FCSS-related changes in network integration and segregation and distinct contributions of large-scale association networks. Together, these findings establish a framework for resolving transient FC state changes and linking their temporal dynamics to task behavior and spontaneous brain-state dynamics.

## RESULTS

### Functional connectivity dissimilarity index (FCD) during rest and task

As depicted in **Fig. 1a,b**, whole-brain dynamic FC was derived from BOLD time series of 300 cortical regions defined by the Schaefer functional atlas[20]. Time-resolved FC was estimated using instantaneous phase coherence (IPC), and eigenvalue decomposition was applied to each IPC matrix to extract its leading eigenvector (V1), representing the dominant whole-brain FC configuration at each time point. Functional connectivity similarity (FCS) was then quantified by comparing V1 patterns averaged within two temporally adjacent windows across multiple window lengths (1–10 TR), with cosine similarities combined using Gaussian weighting to generate a multiscale FCS time series. Rather than the original V1 IPC dissimilarity ($D(t) = 1 - FCS(t)$), squared dissimilarity, $D(t)^2$, was defined as the functional connectivity dissimilarity index (*FCD*) to emphasize pronounced changes in the dominant network configuration. Because its temporal derivative, ($d(D^2)/dt = 2D(dD/dt)$), depends on both the magnitude and rate of change of dissimilarity, this transformation suppresses small background fluctuations while accentuating large, rapidly evolving FC reconfigurations. While correlations of FCD between different ROI datasets (100, 200, and 400 ROIs) were all high, we used the 300 ROI dataset as the representative one whose FCD showing the highest correlation with other dataset.

Resting-state and N-back task fMRI both showed multiple prominent FCD peaks, defined as functional connectivity state switching (FCSS) events based on their characteristic peak morphology and high FCD values (>90th percentile). However, both the FCSS occurrence rate and median FCD were substantially greater during task than rest (**Fig. 1c**). During rest, FCSS events were distributed across time and subjects without consistent temporal alignment, whereas during the N-back task, increases in FCD and FCSS occurrence were strongly time-locked to changes in task state or condition, including rest-to-task and 0-back/2-back switches (**Fig. 1d**). These findings indicate that functional networks undergo stronger and more frequent reconfiguration during task performance, particularly when task demands change and adaptive network reorganization is required. In contrast, resting-state FC generally exhibited smaller configurational changes punctuated by occasional large spontaneous FCSS events.

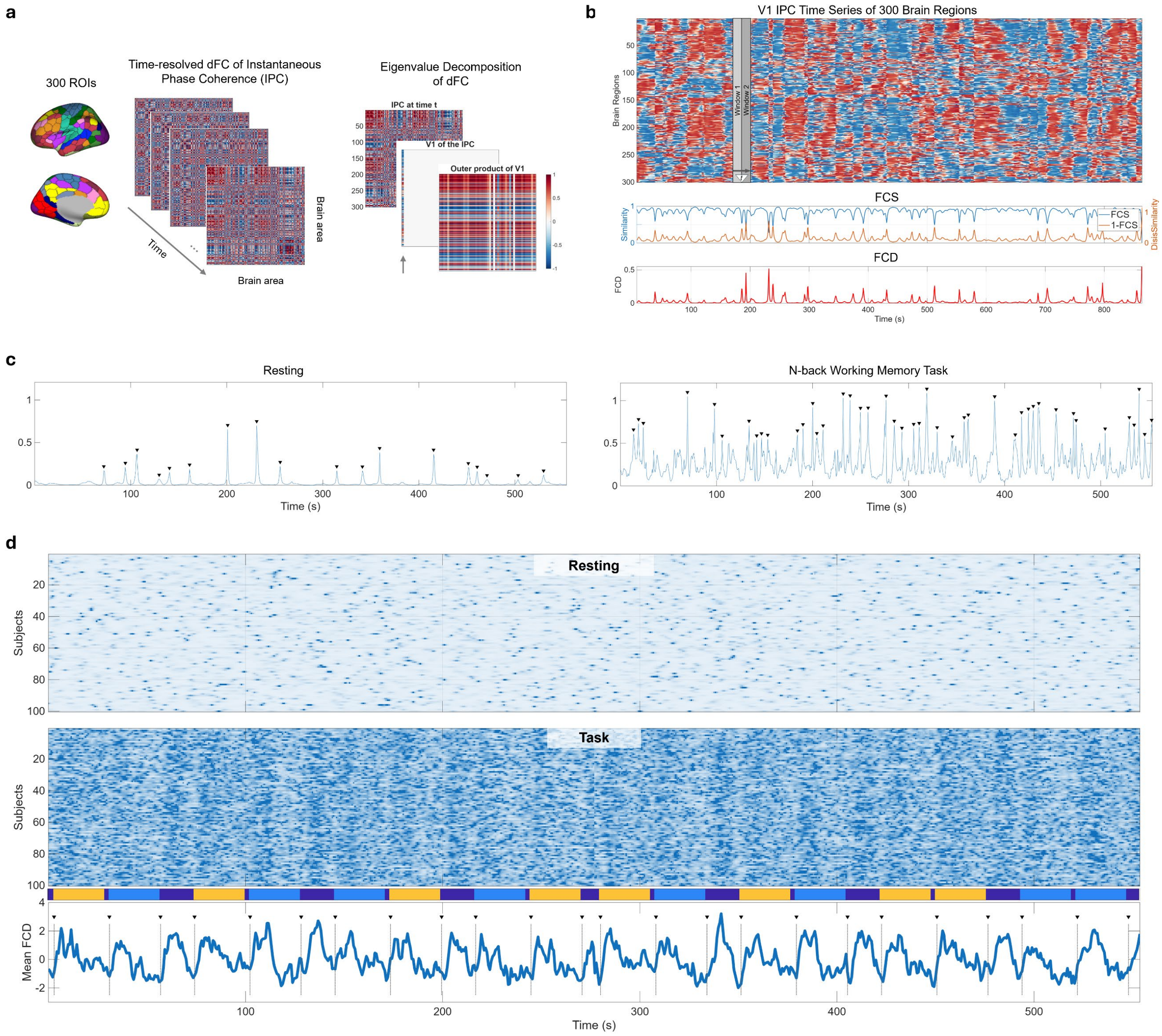

**Figure 1.** FCSS framework and FCD during rest and working memory. (a) Whole-brain BOLD signals were parcellated using the Schaefer atlas, converted to instantaneous phase-coherence matrices, and reduced to the leading eigenvector (V1)

representing the dominant connectivity configuration. (b) Multiscale similarity between temporally adjacent V1 configurations generated the FCS time series; squared dissimilarity defined FCD, with prominent peaks identified as FCSS events. (c) Representative resting-state and N-back FCD time series. (d) Subject-by-time FCD maps during rest and task and the group-mean task FCD time course, illustrating stronger and task-locked reconfiguration during working memory.

## Distinct switching-related FCD dynamics differentially predict task performance

As illustrated in **Fig. 2a**, mean FCD during the N-back task across 100 subjects showed a pronounced temporal correspondence with task-switching events convolved with the canonical hemodynamic response function (HRF). General linear model (GLM) analysis confirmed a robust task-switching effect on FCD ($b$ = 0.43 ± 0.22, $t(99)$ = 19.87, $p < 10^{-36}$, $d$ = 1.99), indicating that changes in task state or condition evoke substantial functional network reconfiguration. Event-related analyses further revealed distinct FCD dynamics for different types of state transitions (**Fig. 2b**). Following task-block onset, FCD increased rapidly and peaked at approximately 4–5 s, closely resembling the temporal profile of the canonical HRF. In contrast, rest-block onset evoked a slower but larger FCD increase, peaking at approximately 6–7 s. These distinct temporal profiles suggest that network reconfiguration depends on the nature of the state transition. Engagement of a new task state produced a rapid reconfiguration accompanying the onset of externally driven processing, whereas transition from task to rest produced a larger and more prolonged reconfiguration, potentially reflecting broader network reorganization following the release from externally imposed task demands.

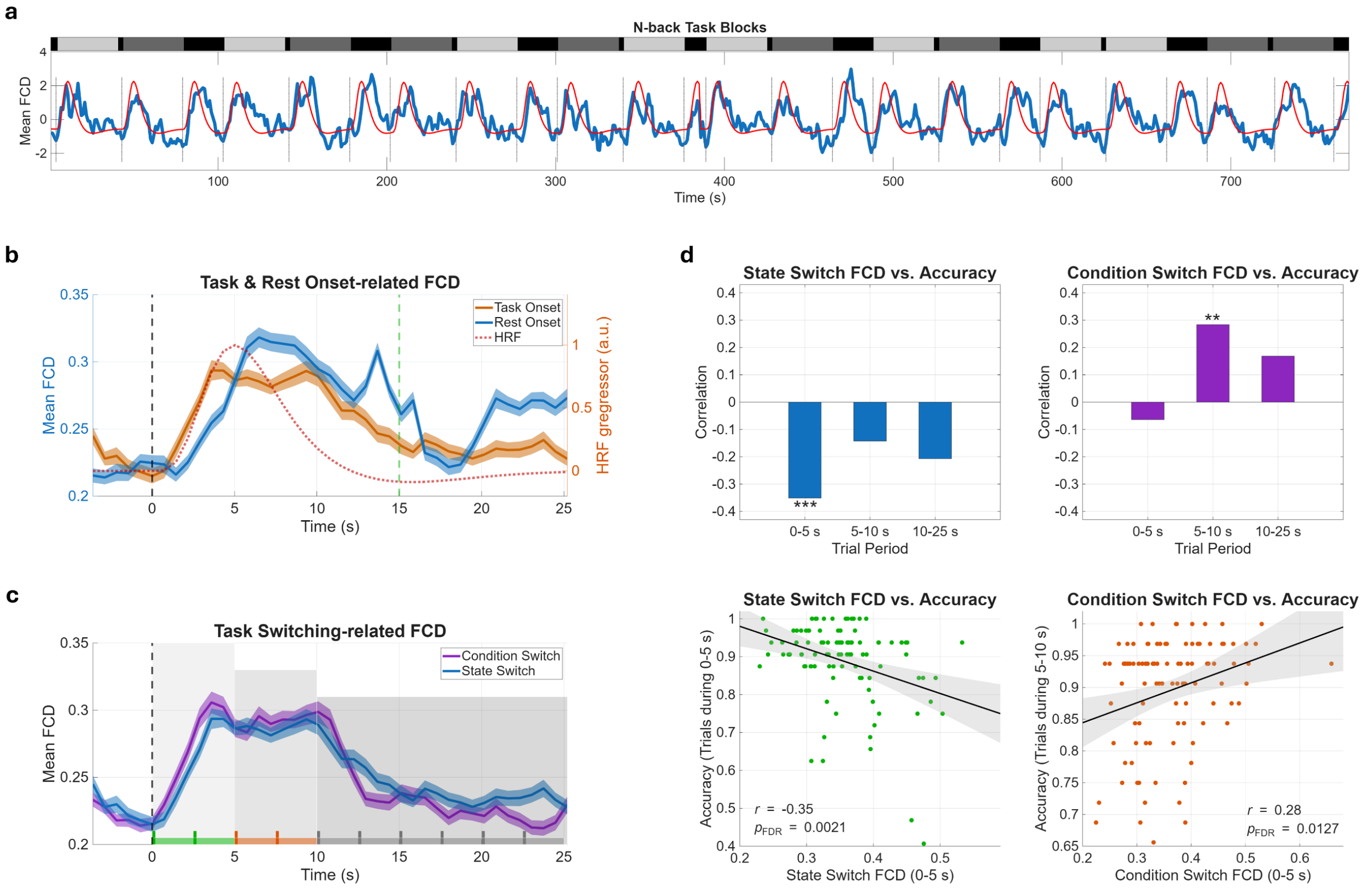

**Figure 2.** Task-related FCD dynamics and behavioral associations. (a) Group-mean N-back FCD and the task-switching event regressor convolved with the canonical hemodynamic response function (HRF). (b) Event-related FCD following task-block and rest-block onsets. Shading denotes uncertainty around the group mean. (c) FCD time courses following rest-to-task state switches and 0-back/2-back condition switches. Colored marks indicate trial epochs used for behavioral analyses. (d)

Correlations between switch-related FCD and response accuracy across successive trial periods, with representative scatterplots for the significant state-switch and condition-switch associations. Reported p values are FDR corrected where indicated.

Further analysis of task-onset FCD revealed distinct behavioral relevance of FCD related to state and condition switches. Rest-to-task state switches and 0-back/2-back condition switches showed broadly similar FCD trajectories, with an early increase peaking within ~5 s, sustained dissimilarity through ~10 s, and a gradual decline thereafter (Fig. 2c). However, condition-switch FCD increased more rapidly, whereas state-switch FCD declined more slowly after ~10 s. Importantly, early state-switch FCD negatively correlated with accuracy for the first two trials occurring within 0–5 s ($r = -0.35$, $p = 0.0003$, $p$FDR = 0.0021), whereas early condition-switch FCD positively correlated with accuracy for the third and fourth trials occurring during 5–10 s ($r = 0.28$, $p = 0.0042$, $p$FDR = 0.0127; Fig. 2d). These opposite, temporally specific relationships suggest that similar FCD increases may reflect different functional processes. Greater rest-to-task reconfiguration may reflect the network adjustment required to establish a task state, such that larger early FCD increases are associated with lower accuracy during the same initial 0–5 s period. In contrast, rapid reconfiguration during condition switching may reflect efficient updating or resetting of an already engaged task network to meet new task demands, which could help explain why larger early FCD increases are associated with better performance during the subsequent 5–10 s period.

**Properties of Spontaneous FCSS**

Task-period FCD was much larger than rest period FCD (median: $t(99) = -95.32$, $p < 10^{-98}$, $d = 9.53$; FCSS peak: $t(99) = -91.26$, $p < 10^{-96}$, $d = -9.13$; **Fig. 3.a**). The FCSS occurrence rate was also much higher during task (4.51 ± 0.37 per minute) than rest (2.13 ± 0.09 per minute; $t(99) = -62.21$, $p < 10^{-80}$, $d = -6.22$). The peaky morphology of FCSS was very similar between task and rest, showing remarkably symmetric rise and fall time course of FCSS switching. However, the task FCSS time course showed a rapid switching within approximately 4 seconds, while rest FCSS showed more gradual switching over around 8 seconds. This switching velocity quantified as the proportion of the areas above FCSS curve (**Fig. 3.b**) was significantly higher in task FCSS than rest FCSS ($t(99) = -39.69$, $p < 10^{-61}$, $d = -3.97$).

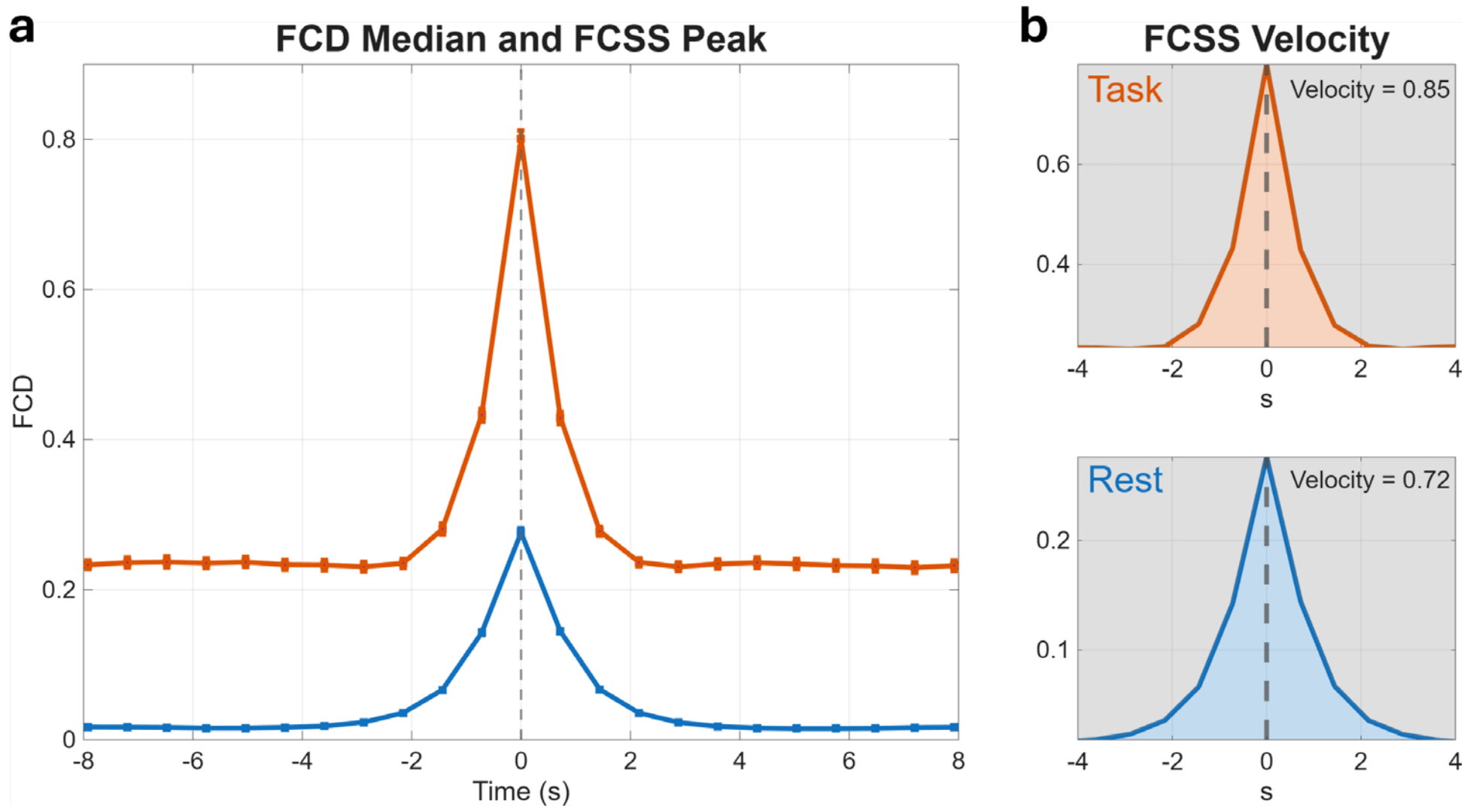

**Figure 3.** Magnitude, morphology and switching velocity of FCSS during task and rest. (a) Event-centered FCD profiles showing the larger FCD baseline and FCSS peak during the N-back task than during rest. (b) Representative task and resting-state FCSS profiles illustrating the switching-velocity index, defined from the area above the normalized FCSS curve relative to its bounding rectangle. Task FCSS was substantially faster than spontaneous resting-state FCSS.

Assessment of the consistency of these measurements of spontaneous FCSS during rest across different runs revealed that the peak value of FCSS was not quite consistent (ICC= 0.18, 95% CI [-0.12, 0.41], $p = 0.1055$) and the occurrence rate showed even worse consistency (ICC = -1.74, 95% CI [-2.73, -0.96]), indicating that within-subject variability exceeded between-subject variability. Additionally, the switching velocity index of FCSS showed a small but statistically significant across-scan consistency (ICC = 0.28, 95% CI [0.02, 0.49], $p = 0.018$), indicating a detectable but weak stable individual differences across scans. However, the median of FCD showed good consistency across different runs (ICC = 0.81, 95% CI [0.74, 0.86], $p < 10^{-27}$). These results indicate that FCSS is state-dependent events rather than stable neural activity reflecting trait of an individual. On the other hand, the median FCD appears to reflect the baseline degree of network state transition, which provides reliable measures of individual differences in network dynamics.

### Brain network properties associated with Spontaneous FCSS

To assess functional brain network properties associated with FCSS, we conducted graph theoretical analysis of functional graphs calculated for three periods of network stability, including stable (FCD ≤ median; 50.3%), switching (FCSS event ± 5 TR; 23.5%), and transition (between stable and switching; 26.2%). We assessed five graph theoretical measures, including a measure of local segregation (clustering coefficient), a measure of global segregation (modularity), a measure of functional integration (global efficiency), a measure of cross-network integration or connector centrality (participation coefficient), and a measure of robustness-related network organization (assortativity). As **Fig. 4b-f** show, clustering coefficient (0.143 ± 0.021, 0.102 ± 0.010, 0.097 ± 0.009), global efficiency (0.219 ± 0.013, 0.194 ± 0.010, 0.192 ± 0.010), and assortativity (0.273 ± 0.078, 0.215 ± 0.060, 0.203 ± 0.057) showed monotonic decreases from stable to switch states. In contrast, participation coefficient showed monotonic increase (0.381 ± 0.090, 0.464 ± 0.062, 0.483 ± 0.053), and modularity was largely comparable across the state, being largest during transition (0.358 ± 0.031, 0.363 ± 0.026, 0.355 ± 0.022).

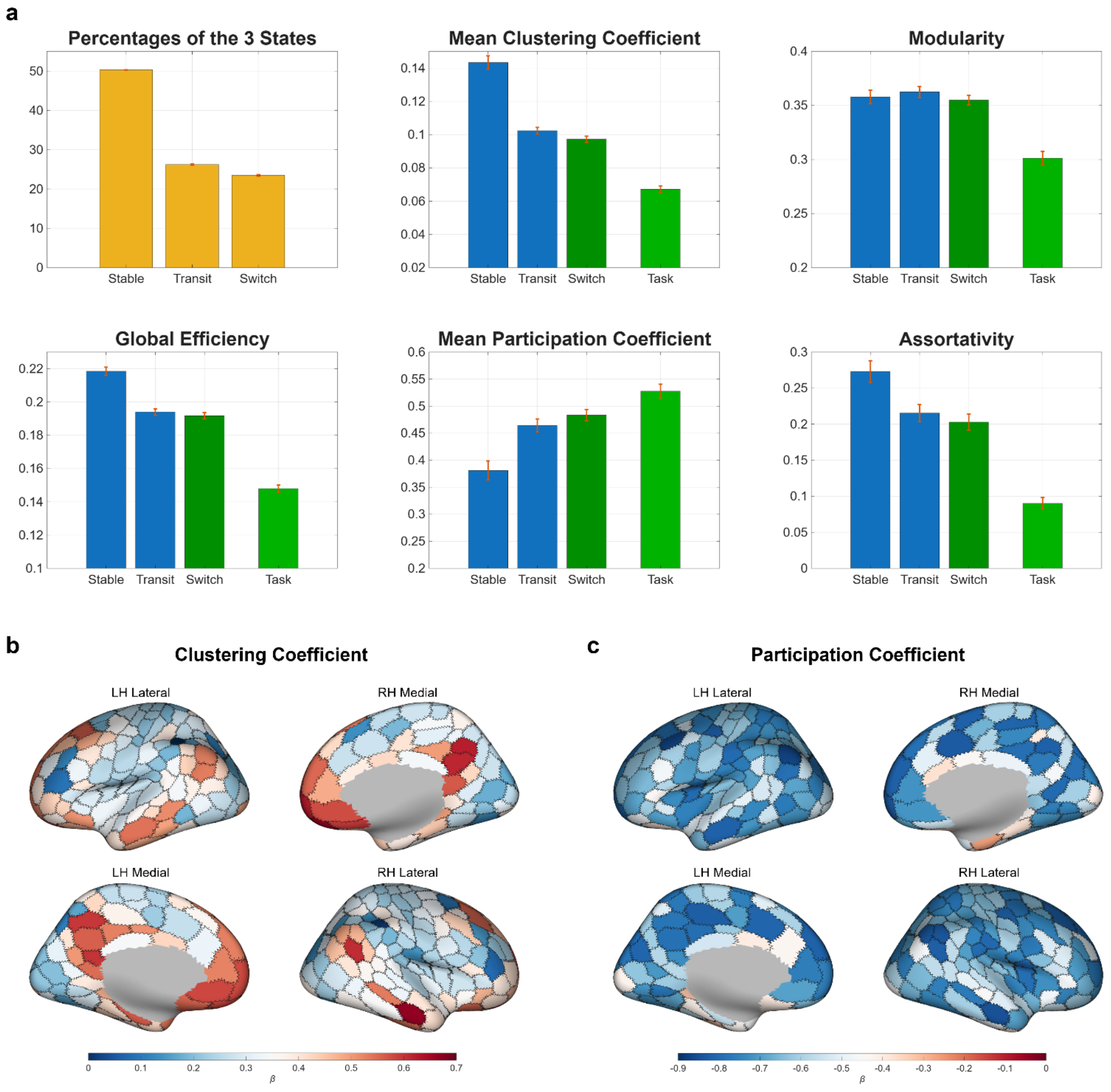

**Figure 4.** Graph-theoretical organization across stable, transition and switching periods. (a) Relative occupancy of the three resting-state periods and group means for clustering coefficient, modularity, global efficiency, participation coefficient and assortativity; N-back task values are shown for comparison. (b,c) Cortical maps of standardized state-effect coefficients for nodal clustering coefficient and participation coefficient across 300 Schaefer parcels. Negative clustering effects indicate progressive loss of local segregation, whereas positive participation effects indicate increasing cross-network connectivity toward FCSS.

Linear mixed effect (LME) modeling confirmed strong negative and positive linear effects of network state on the network measures (mean clustering coefficient: $\beta$ = -0.92, 95% CI [-0.99, -0.84], $t$(298) = -24.19, $p < 10^{-71}$; global efficiency: $\beta$ = -0.82, 95% CI [-0.89, -0.76], $t$(298) = -24.57, $p < 10^{-72}$; participation coefficient: $\beta$ = 0.62, 95% CI [0.52, 0.72], $t$(298) = 12.03, $p < 10^{-26}$; assortativity: $\beta$ = -0.48, 95% CI [-0.55 -0.42], $t$(298) = -15.02, $p < 10^{-37}$), except modularity ($\beta$ = -0.05, 95% CI [-0.14, 0.03], $t$(298) = -1.32, $p$ = 0.1884). Notably, among the three states during rest, graph theoretical measures of switching state were closest to those of the N-back task period functional graph (**Fig. a**). Additional LME analysis on nodal measures of clustering and participation coefficients showed most ROIs showed significant negative and positive FCD effects on their clustering (300 out of 300) and participation coefficient (293 out of 300) after FDR correction. Notably, the majority of the top 10 ROIs showing the strongest FCD effects belong to default mode network, limbic network, and salience network (**Fig. 4.b-c** and **Table 2**).

**Table 1.** Top 10 ROIs showing strong linear effect of network stability on their clustering coefficients and participation coefficient.

| Clustering Coefficient | | | | Participation Coefficient | | | |
|---|---|---|---|---|---|---|---|
| ROIs | $\beta$ | $t$ | FDR-$p$ | ROIs | $\beta$ | $t$ | FDR-$p$ |
| LH DefaultB PFCl 1 | -0.87 | -22.96 | $< 10^{-67}$ | RH DefaultA Temp 1 | 0.67 | 13.69 | $< 10^{-30}$ |
| LH DefaultA IPL 1 | -0.87 | -23.72 | $< 10^{-69}$ | RH LimbicB OFC 4 | 0.66 | 14.26 | $< 10^{-32}$ |
| RH DefaultA Temp 1 | -0.85 | -21.85 | $< 10^{-63}$ | RH LimbicB OFC 5 | 0.63 | 13.79 | $< 10^{-31}$ |
| RH DefaultA IPL 2 | -0.84 | -21.37 | $< 10^{-61}$ | RH DefaultA pCunPCC 3 | 0.62 | 12.84 | $< 10^{-28}$ |
| RH ContB IPL 1 | -0.84 | -20.04 | $< 10^{-57}$ | RH DefaultA pCunPCC 1 | 0.61 | 12.21 | $< 10^{-26}$ |
| RH LimbicB OFC 5 | -0.84 | -20.91 | $< 10^{-59}$ | RH DefaultA IPL 1 | 0.61 | 11.95 | $< 10^{-25}$ |
| RH SalVentAttnA FrMed 1 | -0.83 | -19.80 | $< 10^{-55}$ | LH DefaultA pCunPCC 1 | 0.60 | 13.94 | $< 10^{-31}$ |
| LH SomMotA 2 | -0.83 | -21.78 | $< 10^{-62}$ | LH DefaultA pCunPCC 5 | 0.60 | 13.64 | $< 10^{-30}$ |
| LH SalVentAttnA ParMed 1 | -0.83 | -19.15 | $< 10^{-53}$ | RH LimbicB OFC 3 | 0.59 | 13.35 | $< 10^{-29}$ |
| RH DefaultA IPL 1 | -0.83 | -21.23 | $< 10^{-60}$ | LH DefaultA PFCm 1 | 0.58 | 14.27 | $< 10^{-32}$ |

**Brain regions contributing to spontaneous FCSS**

To identify brain areas whose network contribution showing significant changes prior to spontaneous FCSS events, we analyzed IPC V1 time series of ROIs time-locked to FCSS peak. To minimize noise influence and improve statistical power, we analyzed V1 of 34 ROIs, representing the coarse Schaeffer parcellation patches of visual, somatomotor, dorsal-attention, salience, limbic, control, temporoparietal, and default mode networks[20].

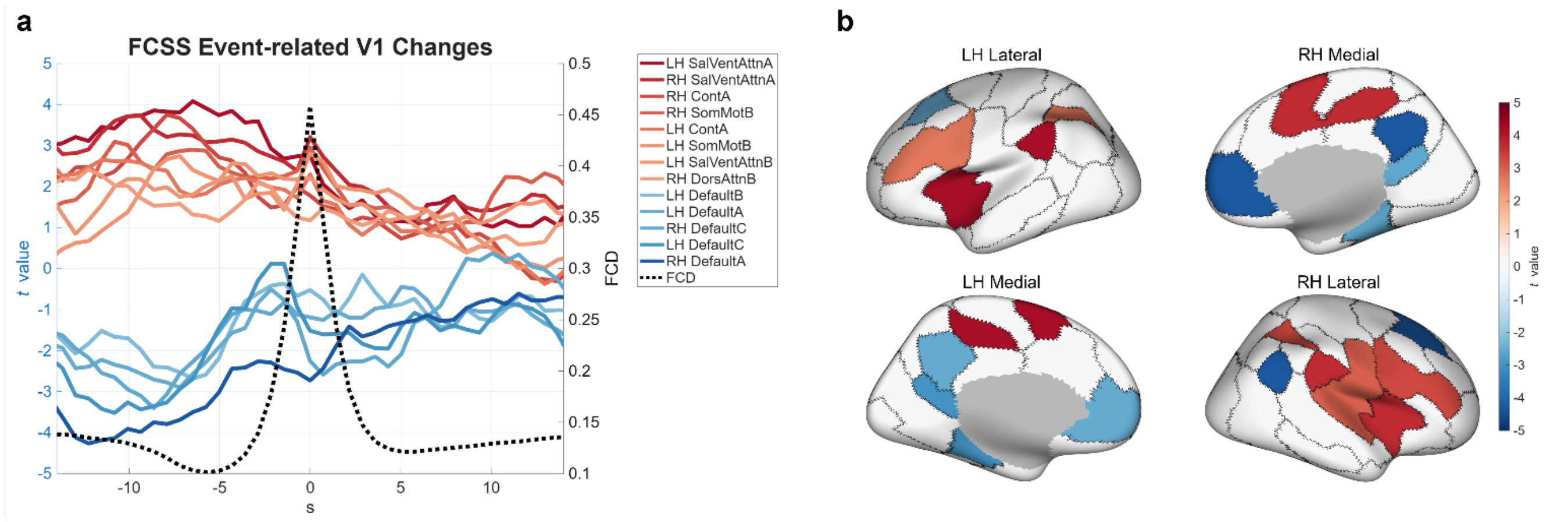

**Figure 5.** Network-specific changes in the dominant phase-coherence configuration around spontaneous FCSS. (a) Event-related V1 trajectories for 34 coarse Schaefer/Yeo-17 hemispheric subnetworks, aligned to the FCSS peak (time 0); the group-

mean FCD profile is overlaid for temporal reference. (b) Cortical maps of the event-related t statistics, showing spatially heterogeneous positive and negative changes in V1 contribution around FCSS.

Event-related V1 analysis revealed systematic, network-specific changes that developed before FCSS peak (-14 to -1 s) rather than a spatially uniform whole-brain shift (Fig. 5). Several default-mode, limbic, salience/ventral-attention and control subdivisions showed among the strongest positive or negative changes in V1 contribution, indicating that spontaneous FCSS reflects coordinated redistribution of the dominant phase-coherence pattern across large-scale association systems. The coexistence of positive and negative regional changes further suggests that the FCD peak reflects reorganization of network configuration rather than a nonspecific global increase in coherence.

## DISCUSSION

The present study introduces FCSS as a continuous, multiscale description of brief whole-brain functional-connectivity reconfiguration and applies it to both externally driven and spontaneous brain-state changes. Four findings are particularly notable. First, FCD and FCSS were markedly enhanced during working memory and strongly time-locked to changes in task state and condition. Second, rest-to-task state switches and 0-back/2-back condition switches produced related but distinguishable FCD trajectories, and their early FCD responses showed opposite, temporally specific relationships with behavioral accuracy. Third, spontaneous resting-state FCSS events were highly state dependent across runs, whereas median FCD showed good across-scan consistency, suggesting that the baseline propensity for network change is more trait-like than the timing or magnitude of individual switching events. Finally, spontaneous FCSS was preceded by broad changes in network topology, including reduced local clustering, global efficiency, and assortativity together with increased participation coefficient, indicating a shift away from the prevailing network configuration as a switch approached. In parallel, event-related V1 analysis showed increasing salience-network involvement and decreasing default-mode-network involvement before FCSS, pointing to a reciprocal change in the relative contributions of these systems during spontaneous network reconfiguration. Together, these findings support a view of FC switching as a transient, coordinated reorganization of whole-brain network configuration with measurable temporal structure rather than an instantaneous jump between predefined discrete states.

FCSS was designed to address a central tradeoff in dFC analysis between temporal resolution and estimation stability. Sliding-window correlation remains widely used but necessarily averages over a temporal interval and can blur brief transitions, while very short windows are vulnerable to sampling variability[1,9,10,40]. Framewise methods improve temporal resolution but emphasize different signal properties: MTD is based on temporal derivatives[11], ETS on instantaneous products of standardized BOLD amplitudes[13], and phase-based approaches on relative phase[12]. FCSS combines a phase-coherence representation with two additional operations. First, the leading eigenvector compresses each IPC matrix into the dominant whole-brain configuration, following the rationale of leading-eigenvector dynamics approaches[18,19]. Second, similarity is evaluated across adjacent windows at multiple scales rather than at a single arbitrarily chosen window length. The resulting FCD therefore represents the temporal trajectory of change in the dominant connectivity configuration. This continuous representation complements state-clustering and hidden-Markov approaches[6,14-16]: it does not require a prespecified number of states and can characterize the rise, peak and resolution of a transition itself.

The task fMRI results provide an external validation of this temporal interpretation. Mean FCD closely followed experimentally defined state and condition changes, and the magnitude of the switching effect was large. Prior work has shown that network flexibility, integration and latent-state occupancy during working memory relate to performance[3-5]; our findings extend this literature by linking the fine temporal evolution of a network transition to behavior at specific positions within a task block. Larger early rest-to-task FCD was associated with lower accuracy during the first 0-5 s, consistent with the possibility that greater reconfiguration reflects a larger adjustment required to establish the task state. In contrast, larger early condition-switch FCD predicted better accuracy during the subsequent 5-10 s, suggesting that rapid reconfiguration of an already engaged task network may support updating to the new rule or working-memory demand. These interpretations are necessarily correlational, but the opposite temporal-behavioral relationships argue against a simple account in which all FCD increases have the same functional meaning.

The close correspondence between task switching and the canonical HRF also requires caution. Because IPC is based on relative phase rather than BOLD amplitude, it is not directly driven by amplitude-dominated cofluctuations associated with task-evoked activation. Nevertheless, a shared task-evoked waveform can align phases across regions, and task-evoked activation can therefore contribute to apparent task-state connectivity even when interregional interaction has not changed[25]. At the same time, we consider the task-locked component potentially part of the biological process of state reconfiguration rather than simply nuisance variance, because the present study concerns how whole-brain functional organization changes with shifts in task demands. Thus, although common evoked responses may contribute to the observed task-related FCSS, the phase-based nature of IPC reduces the direct influence of amplitude changes on the estimated connectivity dynamics. Motion is another important concern in time-resolved connectivity analyses[26,27]; however, associations between FCSS measures and motion parameters were very small and close to zero, suggesting that motion was unlikely to account for the present findings.

The resting-state results distinguish transient FCSS events from more stable individual properties of network dynamics. Recurring dFC states and nonrandom state sequences have been reported across resting-state datasets[6,14,15], and leading-eigenvector analyses have linked spontaneous switching behavior to cognitive performance[18]. In our data, however, FCSS occurrence rate and peak magnitude showed little across-run reliability, whereas median FCD showed good consistency. This pattern suggests that the exact occurrence of a large spontaneous transition is strongly state dependent, while an individual's baseline level of ongoing network configurational change contains a reproducible component. Such a distinction may be useful when considering dFC phenotypes: event counts or individual peaks may be poor trait markers unless estimated from substantially longer recordings, whereas aggregate measures of background network dynamics may be more suitable for individual-difference analyses. The ICC results should therefore be interpreted as properties of specific FCSS metrics rather than as a single global statement about the reliability of dynamic connectivity[36].

Graph-theoretical results further indicate that FCSS is not adequately described as a simple shift from segregation to integration. From stable through transition to switching periods, clustering coefficient fell sharply, indicating loss of locally clustered organization, while participation coefficient increased, indicating broader distribution of connections across network modules[28,32]. At the same time, global efficiency decreased rather than increased[31], implying that greater cross-network participation did not translate into more efficient whole-network shortest-path communication. Assortativity also decreased, consistent with weakening of preferential connectivity among similarly connected nodes and a less robust degree-mixing organization[28,29,33]. Modularity changed little, suggesting that large-scale community structure was comparatively preserved even as local clustering, cross-module participation and hub-related organization changed. Thus, FCSS may represent a transient loosening and redistribution of network architecture: local neighborhoods become less cohesive, connections are spread more broadly across modules, but the network simultaneously becomes less globally efficient and less assortative. The strong nodal effects in default-mode, limbic and salience/control regions are consistent with prior evidence that connector regions and salience-system nodes participate in coordinating interactions among large-scale systems[8,34,35].

The event-related V1 findings provide a complementary network-level perspective on this reorganization. As FCSS approached, the contribution of the salience network to the dominant phase-coherence configuration progressively increased, whereas the contribution of the default-mode network decreased. This reciprocal pattern suggests that switching is not simply a diffuse loss of network stability, but is accompanied by a systematic reweighting of large-scale systems. In conjunction with the graph-theoretical findings, increased salience-network involvement may reflect greater engagement of regions that coordinate communication across functional systems as the existing configuration becomes less locally cohesive and more cross-modular, whereas reduced default-mode contribution may reflect disengagement from the preceding network configuration. Thus, the decreases in clustering, efficiency, and assortativity and the increase in participation coefficient may describe the topological conditions under which switching emerges, while the opposing salience- and default-mode-network dynamics identify a potential systems-level process through which that reconfiguration unfolds. This interpretation is consistent with proposed roles of salience-system regions in detecting behaviorally relevant changes and coordinating transitions among large-scale brain networks[8,34,35], although the present results do not establish a causal role for the salience network in initiating FCSS.

Several limitations should be considered. The temporal precision of FCSS remains bounded by the slow and regionally variable hemodynamic response, and FCD timing should not be interpreted as millisecond-scale neural timing. The present results were obtained from healthy adults in the HCP and from a single working-memory paradigm; replication across independent datasets, tasks and acquisition protocols is required. FCSS detection also depends on choices used for filtering, multiscale weighting and peak definition, although the multiscale formulation reduces reliance on a single window length. Graph measures depend on the construction and thresholding of functional graphs, and robustness across alternative graph definitions should be assessed. Finally, FCSS and the associated V1/graph changes are observational measures of statistical dependence and do not establish causal interactions among regions. Despite these limitations, the convergence of task locking, behavioral associations, spontaneous-event morphology, cross-run reliability patterns, graph-theoretical changes, and event-related V1 dynamics supports FCSS as a physiologically informative framework for characterizing brief whole-brain network reconfiguration. More broadly, the results suggest that brain-state switching has a resolvable temporal anatomy involving coordinated changes in network topology and large-scale system contributions, whose behavioral consequences depend on the states from which and to which the brain transitions.

## METHODS

### Dataset and participants

Functional MRI data were obtained from the WU-Minn Human Connectome Project (HCP), which provides high-temporal-resolution resting-state and task fMRI together with behavioral data[21,23,24]. The primary analyses included randomly selected 100 healthy young adult participants (54 females and 46 males; aged 22-35 years) with complete resting-state and working-memory data. Each participant contributed four resting-state fMRI runs and the HCP working-memory task runs. HCP resting-state data were acquired at TR = 0.72 s with 2-mm isotropic resolution; each resting run contains 1,200 volumes[23]. The HCP working-memory paradigm alternates fixation/rest periods with 0-back and 2-back blocks using visual stimulus categories and is described in detail elsewhere[24].

### Preprocessing and cortical parcellation

We analyzed HCP data after the standard HCP minimal preprocessing pipeline, which includes correction of gradient distortion and head motion, EPI distortion correction, registration to structural images and standard space, and surface-based alignment[22]. Cortical time series were extracted using the Schaefer local-global functional parcellation[20]. The primary analysis used 300 cortical parcels assigned to the Yeo 17-network organization. Robustness of FCD to parcellation resolution was assessed using the 100-, 200- and 400-parcel Schaefer atlases. For coarse event-related localization, the fine parcels were additionally grouped into 34 hemisphere-specific Schaefer/Yeo-17 sub-networks (17 per hemisphere).

**Instantaneous phase coherence and leading-eigenvector representation**

ROI BOLD time series were band-pass filtered (0.008-0.08 Hz) and converted to analytic signals using the Hilbert transform. The instantaneous phase of ROI $i$ at time $t$ was denoted $\varphi_i(t)$. Pairwise instantaneous phase coherence was calculated as $IPC_{ij}(t) = \cos(\varphi_i(t) - \varphi_j(t))$[12]. Thus, each time point was represented by an N × N IPC matrix, where N was the number of cortical parcels. Eigenvalue decomposition was applied to each IPC matrix, and the leading eigenvector V1(t) was retained as a low-dimensional representation of the dominant whole-brain phase-coherence configuration[18,19]. Because eigenvector sign is arbitrary, V1 polarity was aligned across adjacent time points by reversing the sign when the cosine similarity with the preceding V1 was negative. V1 vectors were normalized to unit length before similarity calculations.

**Multiscale functional connectivity similarity and FCD**

To characterize change in the dominant connectivity configuration without fixing a single temporal scale, V1 patterns were compared across two temporally adjacent windows at multiple window lengths. For each window length j = 1-10 TR, V1 vectors were averaged separately within the preceding and following adjacent windows, and cosine similarity was calculated between the two averaged vectors. Similarities across the 10 window lengths were combined using Gaussian weighting that favored shorter windows (sigma = 4 TR) to generate a multiscale functional connectivity similarity (FCS) time series. Dissimilarity was first defined as D($t$) = 1 - FCS($t$). We then defined the functional connectivity dissimilarity index as FCD($t$) = D($t$)$^2$. Squaring preserves the ordering of dissimilarity values while suppressing small background fluctuations and emphasizing large configurational changes. FCD was calculated independently for each run/subject and was the primary continuous measure of network reconfiguration.

**FCSS detection and event-centered morphology**

Functional connectivity state switching (FCSS) events were defined as prominent local maxima in FCD. In the analyses reported here, FCSS peaks were required to exceed the 90th percentile of the FCD distribution and were separated by at least 5 TR. Event-centered FCD epochs were extracted around each peak and averaged within subject before group-level analysis. FCSS peak magnitude was defined as FCD at the event center. Occurrence rate was expressed as events per minute. Switching velocity was quantified from the normalized event-centered FCSS profile as the proportion of the bounding-rectangle area lying above the FCSS curve (equivalently, 1 minus the normalized area under the curve), such that sharper and more rapidly evolving peaks produced larger values.

**Working-memory task-switching analyses**

Task timing was obtained from the HCP working-memory event files[24]. Three classes of task changes were distinguished: rest-to-task state switches, task-to-rest state switches, and condition switches between 0-back and 2-back blocks. A task-switching event vector was convolved with the canonical hemodynamic response function (HRF) and entered into a subject-level general linear model predicting FCD. The resulting unstandardized coefficient b quantified the magnitude of task-switch-related FCD modulation for each participant; group significance was tested across participants. Event-related FCD was also averaged separately around task-block onset, rest-block onset, state switches and condition switches to compare their temporal profiles.

**Behavioral analyses**

To test whether early network reconfiguration was related to behavioral adaptation, event-related FCD was summarized over successive trial periods following task onset/switching. Accuracy was calculated for trials occurring within the corresponding temporal intervals. Primary analyses focused on the first two trials (approximately 0-5 s), the third and fourth trials (approximately 5-10 s), and the later portion of the block (approximately 10-25 s). Pearson correlations related switch-associated FCD measures to accuracy across subjects. State-switch and condition-switch analyses were performed separately because the former establishes a task state from rest whereas the latter updates the rule within an already engaged task context. Multiple behavioral correlations were controlled using the Benjamini-Hochberg false discovery rate (FDR) procedure[37].

**Resting-state FCSS properties and across-run consistency**

For each resting-state run and participant, we calculated median FCD, FCSS peak magnitude, FCSS occurrence rate and switching velocity. Task and rest values were compared within participants using paired tests, with Cohen's *d* reported as an effect-size estimate. Across-run stability of resting-state metrics was evaluated using consistency intraclass correlation coefficients for the average of four runs, ICC(C,4), together with 95% confidence intervals[36]. ICCs were interpreted as properties of each specific metric; positive statistical significance was distinguished from the magnitude of reliability. The primary purpose of this analysis was to determine whether the timing/magnitude of individual FCSS events or more aggregate FCD properties carry stable individual-difference information.

**Graph-theoretical analyses**

Resting-state time points were classified into three ordered FCSS periods: stable (FCD less than or equal to the subject/run median), switching (within +/-5 TR of an FCSS peak), and transition (remaining time points between stable and switching periods). These definitions yielded approximately 50.3%, 23.5% and 26.2% of resting time points, respectively. For each participant and period, a functional graph was constructed from the corresponding data by computing Pearson correlation matrix and zeroing diagonal and negative values. We quantified mean clustering coefficient (local segregation), modularity (global community segregation), global efficiency (whole-network integration/communication efficiency), participation coefficient (cross-network integration/connector centrality) and assortativity (degree-mixing/robustness-related organization)[28-33]. Participation coefficients used the Schaefer/Yeo network assignment as the module definition. Nodal clustering and participation coefficients were additionally calculated for each of the 300 parcels.

**Linear mixed-effects modeling of ordered FCSS periods**

To identify graph properties that changed progressively across stable, transition and switching periods, FCSS period was encoded as an ordered numeric predictor (Stable = 1, Transition = 2, Switching = 3). Because graph measures have different numerical scales, each dependent graph measure was standardized before modeling. Linear mixed-effects models included state as a fixed effect and participant as a random intercept: GraphMeasure ~ State + (1|Participant). The standardized state coefficient beta therefore represents the change in the graph measure, in standard-deviation units, for each one-step progression toward switching. The same model was applied separately to nodal clustering and participation coefficients. FDR correction was applied across the 300 nodal tests for each measure[37]. Pairwise within-subject contrasts among the three periods were treated as secondary analyses.

### Event-related V1 analysis

To identify cortical systems whose contribution to the dominant phase-coherence mode changed around spontaneous FCSS, V1 time series were computed for BOLD time series of the 34 coarse hemisphere-specific Schaefer/Yeo-17 subnetworks and epoched relative to FCSS peaks. Event-related V1 trajectories were averaged within participant and tested across participants at the prespecified time point/interval around the FCSS peak. The resulting *t* statistics were projected onto the cortical surface to visualize subnetworks showing positive versus negative changes in contribution to the dominant connectivity configuration.

### Statistical analysis

Unless otherwise specified, analyses were conducted across N = 100 participants. Two-sided paired t tests were used for within-subject comparisons, and Pearson correlations were used for continuous brain-behavior associations. Linear mixed-effects models accounted for repeated measurements across FCSS periods. Standardized beta coefficients, t statistics, confidence intervals and p values were reported for mixed-effects analyses. Multiple comparisons were controlled by the Benjamini-Hochberg FDR procedure where families of related tests were performed[37]. Statistical significance was evaluated at alpha = 0.05. Analyses were implemented in MATLAB R2025b.